\documentclass[twocolumn]{aa}
\usepackage{graphicx}
\begin{document}
   \title{In-flight calibrations of IBIS/PICsIT\thanks{Based on observations with
INTEGRAL, an ESA project with instruments and science data centre funded by ESA member
states (especially the PI countries: Denmark, France, Germany, Italy, Switzerland, Spain),
Czech Republic and Poland, and with the participation of Russia and the USA.}}

   \author{G. Malaguti\inst{1}, A. Bazzano\inst{2}, A.J. Bird\inst{3}, G. Di Cocco\inst{1}, L. Foschini\inst{1}, P. Laurent\inst{4}, A.~Segreto\inst{5}, J.B.~Stephen\inst{1}, P.~Ubertini\inst{2}
          }

   \offprints{G. Malaguti  - \email{malaguti@bo.iasf.cnr.it}
}

   \institute{$^1$Istituto di Astrofisica Spaziale e Fisica Cosmica - IASF/CNR - Sezione di Bologna\\
$^2$Istituto di Astrofisica Spaziale e Fisica Cosmica - IASF/CNR - Roma\\
$^3$School of Physics and Astronomy, University of Southampton, Highfield, Southampton, SO17 1BJ, UK\\
$^4$CEA, Saclay, France \\
$^5$Istituto di Astrofisica Spaziale e Fisica Cosmica - IASF/CNR - Sezione di Palermo\\
}

   \date{Received ; accepted }

   \abstract{PICsIT (Pixellated Imaging CaeSium Iodide Telescope) is the high energy detector of the IBIS telescope on-board the INTEGRAL satellite. It consists of 4096 independent detection units, $\sim$0.7 cm$^2$ in cross-section, operating in the energy range between 175 keV and 10 MeV. The intrinsically low signal to noise ratio in the gamma-ray astronomy domain implies very long observations, lasting $10^5-10^6$ s. Moreover, the image formation principle on which PICsIT works is that of coded imaging in which the entire detection plane contributes to each decoded sky pixel. For these two main reasons, the monitoring, and possible correction, of the spatial and temporal non-uniformity of pixel performances, expecially in terms of gain and energy resolution, is of paramount importance. The IBIS on-board $^{22}$Na calibration source allows the calibration of each pixel at an accuracy of $<0.5\%$ by integrating the data from a few revolutions at constant temperature. The two calibration lines, at 511 and 1275 keV, allow also the measurement and monitoring of the PICsIT energy resolution which proves to be very stable at $\sim$19\% and $\sim$9\% (FWHM) respectively, and consistent with the values expected analytical predictions checked against pre-launch tests.
   \keywords{gamma-ray telescopes --
             imaging detectors --
             gamma-ray astronomy
               }
   }
\authorrunning{Malaguti et al.}
\titlerunning{In-flight IBIS/PICsIT calibrations}

   \maketitle
%

\section{Introduction}

PICsIT (Di Cocco et al. 2003) is the high energy detection plane of the IBIS (Ubertini et al. 2003) telescope. 
It consists of 8 independent rectangular modules arranged in a $2\times4$ array, 
each containing 512 ($16\times32$) CsI(Tl) scintillator crystals coupled with 
silicon photodiodes (Si-PD) for fluorescence light collection.
The signals from the 4096 independent detection units (pixels) are first equalized by means of an onboard look-up table (LUT), to account for pixel performance disuniformities, and then transformed into the corresponding energy value.

PICsIT detects the gamma-ray shadow projected by the coded mask placed $\sim$3.3 m above it, and the sky image is produced by a deconvolution of this shadowgram with the mask pattern (see Goldwurm et al. 2003, for details regarding IBIS data analysis). By using this technique, every pixel of the detection plane contributes to the entire image of IBIS field of view, thus making the minimization of pixel-by-pixel disuniformities a key aspect of the PICsIT calibration.

Because of telemetry budget limitation at satellite level, PICsIT cannot work in photon-photon mode. In the standard science mode (see Di Cocco et al. 2003 for details about PICsIT operative modes), PICsIT events are summed together on-board to form histograms, which are then sent to ground. In this way the information on single raw events is ``lost'' after the on-board correction for pixel gain non-uniformity, which becomes a key aspect of data reduction and analysis.

The dominant physical factors affecting pixel performances and the channel-to-energy relationship, which can in principle cause disuniformities among pixels are: crystal scintillation light yield and uniformity, efficiency of light collection at PD, CsI-PD optical coupling, PD spectral response and quantum efficiency, pixel gain and linearity. 
For simplicity, in what follows, we will indicate as {\em gain} the aggregate result of all these factors.

%
   \begin{figure*}
   \centering
   \includegraphics[height=7cm,width=18cm]{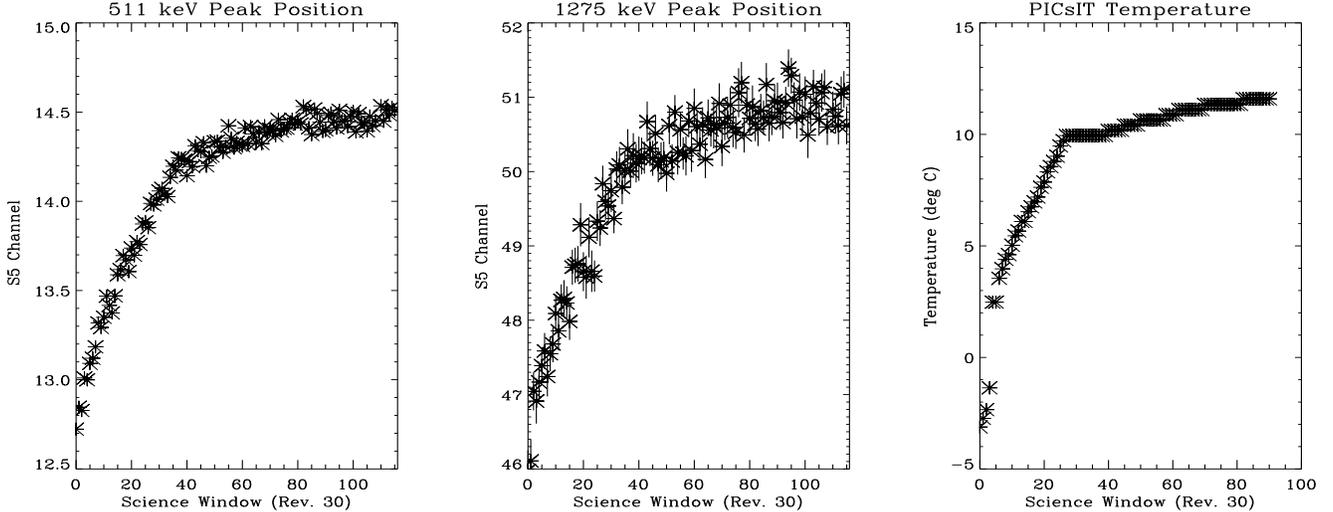}
   \vspace*{0.5cm}
   \caption{Average (S5 spectra of all PICsIT pixels summed together) 511 keV and 1275 keV 
(left and centre panels) peak position evolution along the INTEGRAL orbit during revolution 30 (1 Science Window = 2000 s), compared with the PICsIT temperature trend from one of the 32 sensors onboard (the error on the temperature is $\sim$10\%). }
    \label{Peak_vs_T}
    \end{figure*}

The intrinsic weakness of the gamma-ray flux coming from cosmic sources with respect to the background (the signal to noise ratio, SNR, for intense sources is $\sim$few \%), implies the necessity of long ($10^5\div10^6$ s) observing times.
Given these long integration times, the gain of each PICsIT pixel must be known with very high accuracy in order to minimize the effect of systematic errors. 

\section{On-board gain determination and monitoring}

Pre-launch measurements were used to create LUTs describing the pixel characteristics. Subsequent gain variations are expected during the operative lifetime of PICsIT over different time scales, due to different causes. A {\em prompt} gain variation can be expected following the thermal and mechanical shocks of the launch which may determine a variation in the light collection efficiency (crystal wrapping and/or PD optical coupling). A {\em medium} time-scale variation is associated (see section 2) with temperature gradients along the INTEGRAL orbit, while a {\em long} time-scale variation can be expected from ageing effects (e.g. of the CsI crystal fluorescent yield).

These gain variations can be classified into two broad classes: 
a) uniform: the gain of each pixel drifts by the same amount in the same direction, and b) random: pixel-to-pixel differences. 
For class b, the spatial frequency (which gives the distribution of pixel-to-pixel gain variations across the detection plane) is maximum if the gain drift of each pixel is completely independent from the others, while it can have some kind of structure in the presence of temperature gradients.

In the conservative hypothesis of pixel-by-pixel independent gain variations, the effect of systematics introduced by the gain disuniformity increases with the total number of counts and therefore with observing time. In order to limit the sensitivity degradation to around 10-15\%, an accuracy of $\sim$1\% in the gain determination for each pixel is needed for a $10^5$ s observation, which becomes $\sim0.5$\% for a $10^6$ observation (Malaguti et al. 1999).

An on-board calibration unit (OBCU) is employed by the IBIS system to monitor the performances of the two detection planes primarily in terms of linearity and energy resolution (Bird et al. 2003). The OBCU is based on a $0.4\mu$Ci (at launch date, half life = 2.6 yrs) $^{22}$Na radioactive source coupled with a modified IBIS anticoincidence BGO crystal module (Quadrini et al. 2003). The unit is placed at 220 cm above the upper detection plane to maximize the illumination uniformity. $^{22}$Na decays, with a branching ratio of 90\%, into three photons simultaneously, one at 1.275 MeV and two at 0.511 MeV which are emitted back-to-back.
The signal from any PICsIT pixel coincident (within a 3$\mu$s time window) with the detection by the BGO of a 511 keV photon from the the OBCU is 
tagged by the on-board data handling. The tagged events are not corrected by the onboard PICsIT LUTs, but are put in a dedicated telemetry data stream. These PICsIT-OBCU coincidence data are first accumulated onboard with an integration time set to 1800 s to form a $64\times64\times64$ cubic array containing a 64 channel spectrum for each of the 4096 pixels, and then sent to ground as part of the scientific housekeeping data.

%
   \begin{figure}
   \centering
   \includegraphics[width=9cm]{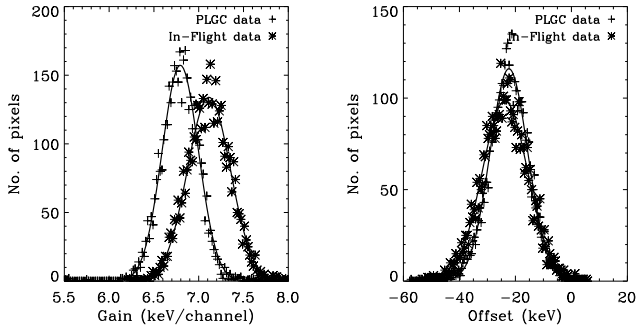}
   \caption{PICsIT pixel-by-pixel gain (left panel) and offset (right panel) distributions obtained during PLGC tests (+) and accumulating OBCU data at constant temperature from rev. 39 to rev. 45 (*).}
   \label{gaofs}
   \end{figure}

\section{Gain temperature dependence}

%
   \begin{figure*}
   \centering
   \hspace{-0.5cm}
   \includegraphics[scale=0.53]{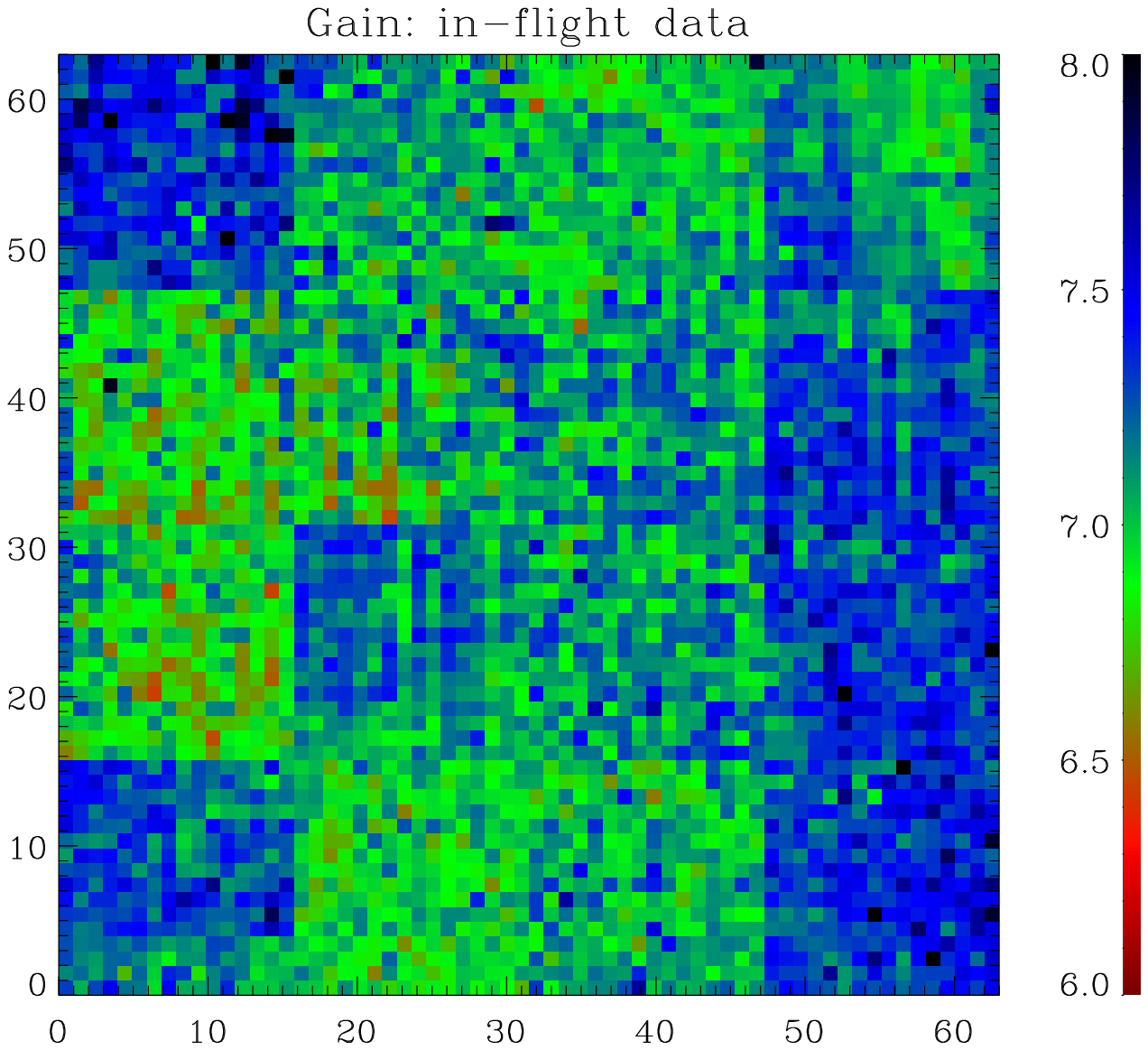}\hspace{-2.3cm}
   \includegraphics[scale=0.53]{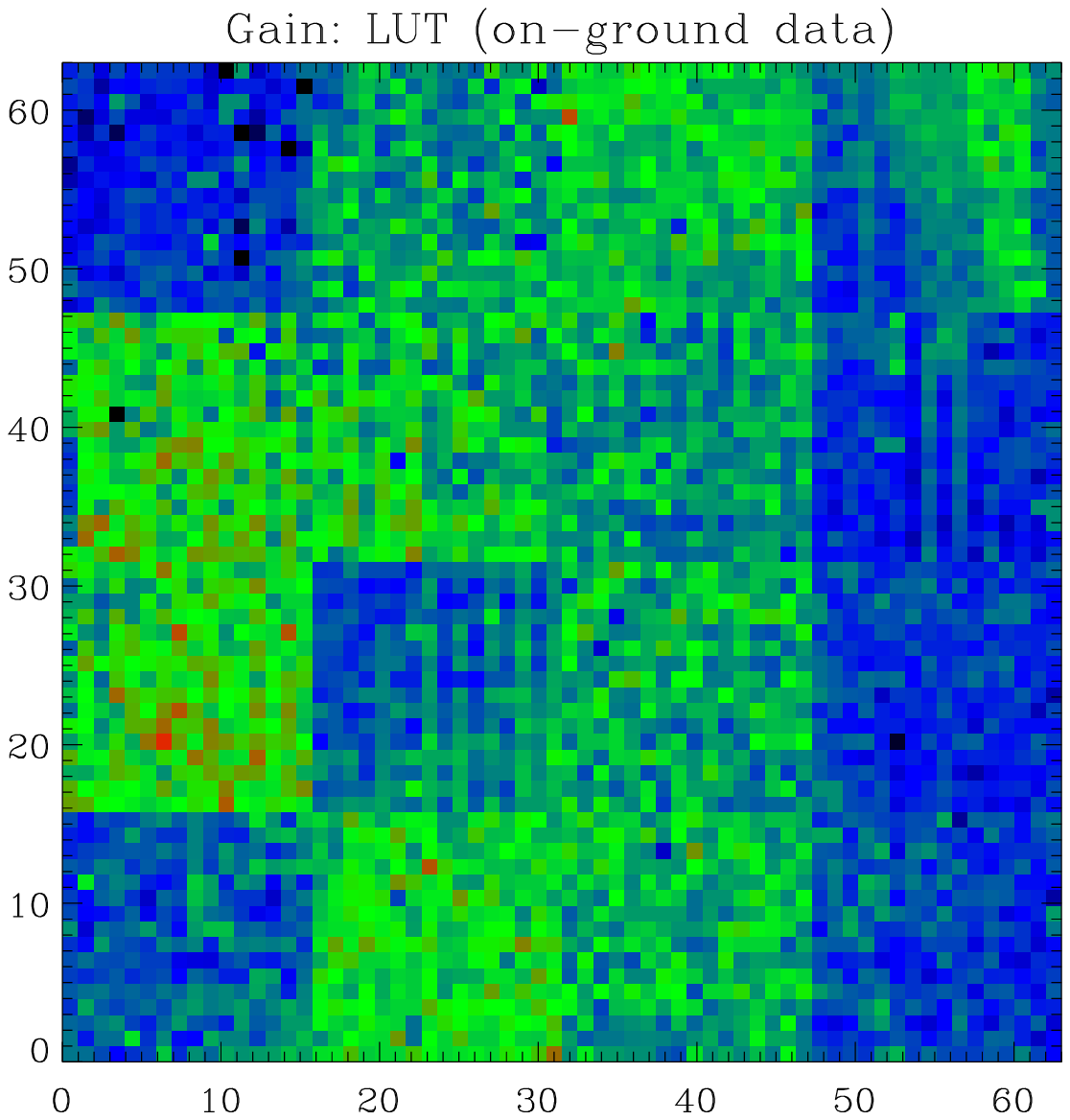}\hspace{-3.2cm}
   \includegraphics[scale=0.53]{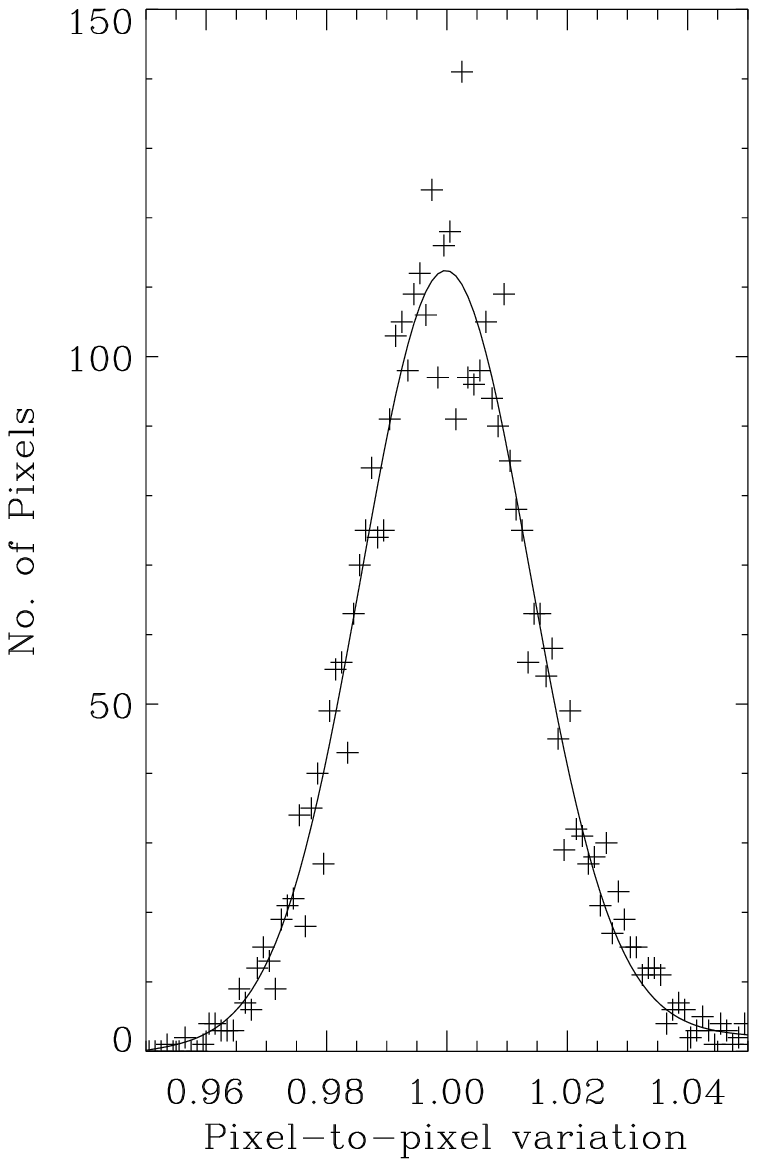}
   \caption{PICsIT pixel-by-pixel gain distributions (the color scale is in keV/channel, see equation 1) obtained from in-flight calibration data (left panel) compared with the onground values used to create the presently on-board LUTs (central panel). The pixel-by-pixel fractional difference is also shown (right panel).}
   \label{LUTs}
   \end{figure*}
%

The dependence between gain and temperature for PICsIT pixels has been measured during the INTEGRAL pre-launch thermal tests performed in May 2002, and a coefficient of $\simeq0.3\%/^\circ$C has been found between $-5$ and 25 $^\circ$C.

By means of the 32 (4 for each module) temperature sensors placed on PICsIT, and using the OBCU-PICsIT spectra, the gain variation 
with temperature has been re-measured onboard during the first months of operation of PICsIT in orbit.

Figure~\ref{Peak_vs_T} shows the variation of the 511 and 1275 keV peak centroid positions from the IBIS CU data obtained during revolution 30\footnote{PICsIT was switched off during belt transits only up to revolution 37; from revolution 38 onwards, PICsIT is always switched on, so that the temperature profile along the orbit is nearly constant, and the only significant variations are due to the solar aspect of the satellite.} compared with the PICsIT temperature evolution during the same orbit.
The errors in the 1275 keV line peak positions are larger because of the lower statistics for two main reasons: a 1275 keV photon is emitted by $^{22}$Na approximately every two 511 keV photons, and the OBCU tagging efficiency is slightly lower at this energy (Bird et al. 2003).

It is also important to note that the $y$ axis of Figure \ref{Peak_vs_T} refers to the OBCU binning table which is not linear. In fact, in order to optimize the energy sampling in the regions of the two lines within the required limit of 64 channels imposed by telemetry budget, channel 0 and channel 32 contain 48 and 40 raw channels, respectively, while the remaining channels contain two raw channels each.
The linearized peak position evolutions coupled with the temperature variation along the orbit shown in the right panel of Figure~\ref{Peak_vs_T} give a gain-temperature dependence of $0.33 \%/^\circ$C.
Taking into account the expected gain variation of $0.3 \%/^\circ$C measured during on-ground thermal tests, the profiles shown in Figure 1 clearly indicate that the PICsIT temperature gradient along INTEGRAL orbit is the dominant factor in the observed gain variation.

This gives the possibility to perform PICsIT pixels gain monitoring 
by producing a pseudo-temperature map, measuring the position of only the 511 keV line from the OBCU, thus allowing the detection of possible gain variations on a shorter time scale.

\section{On-Board Gain Determination and Pixel Equalization}

The channel to energy relationship is normally expressed by the following equation:
\begin{equation}
{\rm Energy} = {\it Offset} + {\rm Channel} \times {\it Gain},
\end{equation}
where, if the {\em Energy} is expressed in keV, then the {\em Gain} is in keV/Channel, and the {\em Offset} in keV. One of the main aims of the on-board calibration unit is the determination and continuous monitoring of gain and offset for each pixel, at the required accuracy level (see section 2).

Given the observed dependence upon the temperature, the first determination of the inflight gain and offset for each pixel has been performed using the OBCU data acquired during the long calibration exposures dedicated to the observation of the Crab, from revolution 39 to revolution 45. The constant pointing of the telescope during this period, has ensured that possible temperature fluctuations due to variation of the solar aspect are avoided.

Figure~\ref{gaofs} shows the distributions of gain and offset for all PICsIT pixels, comparing the inflight data with the results of the measurements performed during INTEGRAL PayLoad Ground Calibration (PLGC). 
The PLGC and in-flight gain distributions have a mean value of 6.8 and 7.1 keV/channel, respectively.
The misalignement of the two gain distributions is fully explained by the different operating temperatures: $\simeq5^\circ$C during Crab observations, and $\simeq20^\circ$C during PayLoad Ground Calibrations (PLGC), thus confirming the $0.3\%/^\circ$C relation. The offset distributions do not show significant dependence upon the temperature ($<$0.2 channels), thus opening the possibility to calibrate using only one line, decreasing in this way the integration time needed.

The long stable Crab pointing, lasting 7 revolutions ($\sim$3 weeks) has allowed the comparison of pre-launch gains with in-flight values in order to verify the presence and magnitudes of these variations. 
Figure~\ref{LUTs} shows the gain maps from inflight (left panel) and on-ground (centre panel) data, together with the pixel-by-pixel fractional difference distribution (right panel). 
The map of the gains measured in-flight maintains the same semimodule and pixel structures already present in the pre-flight results. These are due to the fact that, in order to minimize their internal disuniformity, every semi-module has been populated with pixels coming from manufacturing lots having the same average performances.
The similarity of these structures between pre-launch and in-flight data (Figure~\ref{LUTs}, left and centre panels), indicates the absence of significant post-launch variations, and the right panel of Figure~\ref{LUTs} quantifies this pixel stability within 3.2\% (FWHM).

\section{Energy resolution and detector stability}

\begin{figure}
   \centering
   \includegraphics[width=9cm]{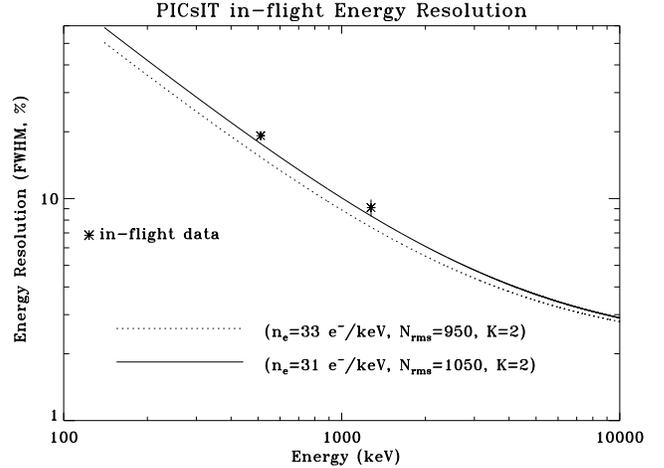}
   \caption{PICsIT energy resolution vs. energy: in-flight measurement are shown together with the theoretical profiles calculated for two set of parameters measured during ground calibrations.}
   \label{DEvE1}
   \end{figure}

The OBCU gives also the opportunity to check the long term stability of the PICsIT energy resolution, to check against possible temperature dependence (orbit time-scale), and/or degradations due to ageing effects or after-launch damages (longer time-scale). The energy resolution of PICsIT can be defined in terms of the signal produced by an interaction and the associated noise, and depends upon the number, $n_{\rm e}$, of electrons created by an energy deposit $E$, and the PD electronic noise, which is normally expressed by means of the standard deviation of number of electrons, $N_{\rm rms}$. The energy resolution of a single CsI-PD pixel is then:
\begin{equation}
\left(\frac{\Delta E}{E}\right)_{\rm FWHM} = 2.35\frac{\sqrt{n_{\rm e}E+N_{\rm rms}^2}}{n_{\rm e}E} + K ,
\label{eq_DE}
\end{equation}
where $K$ is an additional term that accounts for crystal non-uniformity mainly due to variation in the thallium concentration. 
Figure \ref{DEvE1} shows the energy resolution measured on-board at 511 and 1275 keV, compared with the theoretical profile given by equation \ref{eq_DE} with different sets of parameters measured during ground calibrations. 
%
The slight worsening with respect to pre-launch theoretical values is to be ascribed to variations in the light output of the CsI-PD due to changes in the optical contact (epoxy to silicon) after thermal tests. Moreover, the electronic noise was calculated during the tests of the engineering and qualification models, and refers to ASIC only, while a $\sim$10\% increase is expected when going to the complete detector electronic chain.

   \begin{figure}
   \centering
   \includegraphics[width=9cm]{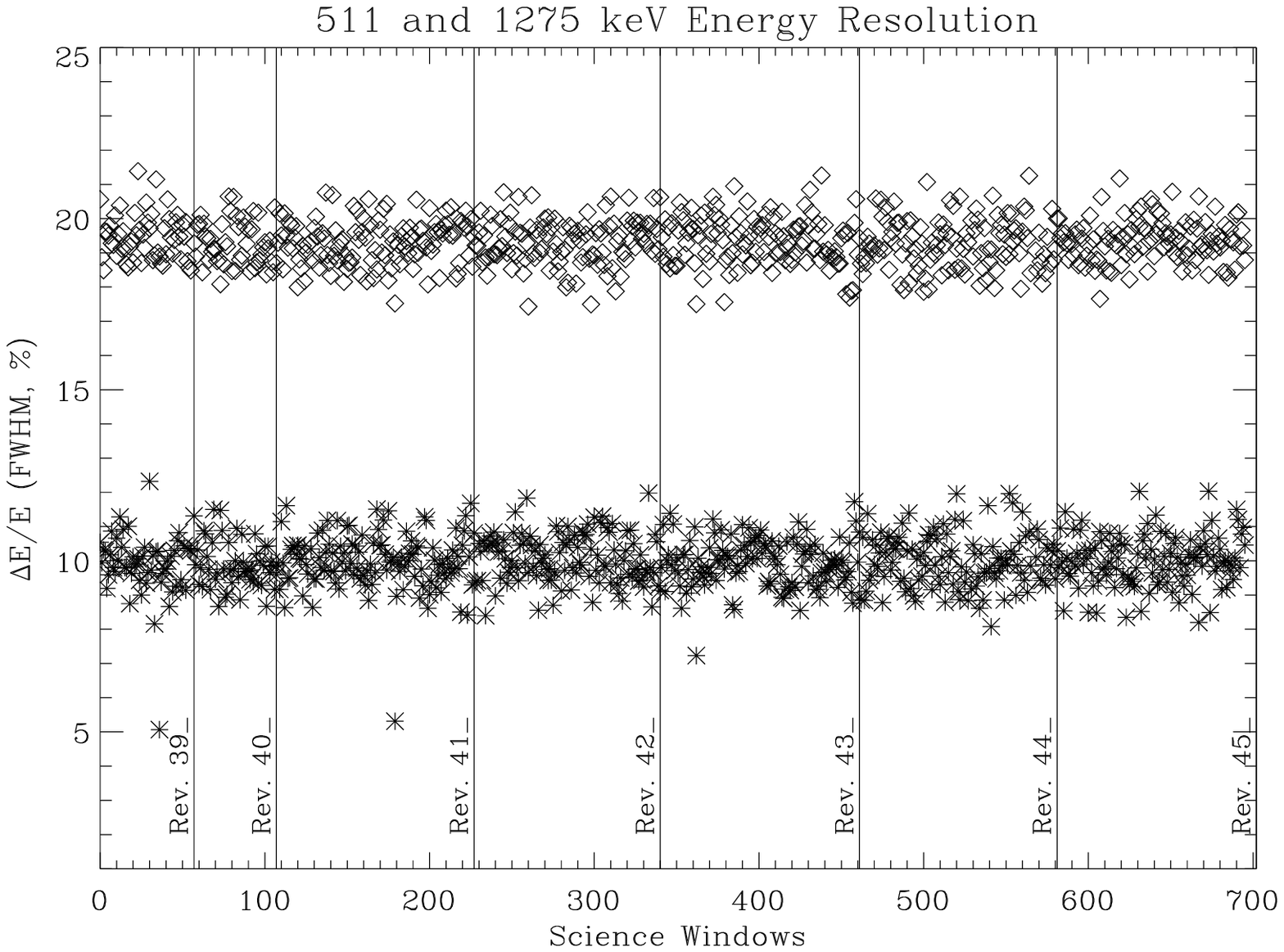}
   \caption{PICsIT energy resolution at 511 and 1275 keV measured during the Crab pointing from revolution 39 to 45. Every data point indicates one science window.}
   \label{DEvE2}
   \end{figure}
%
The long-term stability of the PICsIT energy resolution has also been measured during the first 8 months of INTEGRAL.
Figure \ref{DEvE2} shows the energy resolution, corrected for offset, at the two OBCU line energies measured for a period of $\sim$3 weeks from revolution 39 to revolution 45. The data indicate the stability of the instrument, together with the absence of degradation or changes.

\section{Conclusions}

In the first months of operation of INTEGRAL, the IBIS OBCU has demonstrated an overall very good and stable behaviour of the PICsIT pixels in terms of gain and energy resolution, together with the absence of significant variations with respect to pre-launch values.

In-orbit gain variations are dominated by temperature changes, so that PICsIT temperature values can be used to monitor gain variations on short timescales ($\leq 1$ revolution). Moreover, the single pixel gain stability can be checked at the required accuracy, within medium timescales ($\sim1\div2$ revolutions),  by means of the OBCU 511 keV line only.

On longer timescales, the on-board LUTs are periodically recalculated with OBCU data to verify the presence of ageing effects. 

The measurements reported in this paper will be used to upload the first in-flight LUTs which take into account the effects due to the mechanical and thermal shocks of the launch, and the fact that during the long calibration pointings, the temperature was 5$^\circ$C instead of 20$^\circ$ as during the on-ground calibrations.

\begin{acknowledgements}

This work has been partially funded by the Italian Space Agency (ASI).
The italian participation to the INTEGRAL/IBIS project is financed by the Italian Space Agency (ASI). Fruitful collaboration with LABEN S.p.A., Alenia Spazio, and the use of the Calibration Facility at ESA/ESTEC is kindly acknowledged.
LF acknowledges the kind hospitality of the INTEGRAL Science Data Centre during part of this work.
AJB acknowledges funding by PPARC grant GR/2002/00446.
\end{acknowledgements}

\end{document}